\documentclass[runningheads]{llncs}

\usepackage{graphicx}
\usepackage{csquotes}
\usepackage{amsfonts}
\usepackage{hyperref}
\usepackage{makeidx}
\usepackage{listings}
\usepackage{float}
\usepackage{mpshaskell}
\usepackage{subcaption}

\newcommand{\seqset}{\textsc{FunSeqSet}}
\newcommand{\link}{\haskell{link}}

\newcommand{\FT}{\haskell{FT}}

\begin{document}

\title{FunSeqSet: Towards a Purely Functional Data Structure for the Linearisation Case of Dynamic Trees Problem \thanks{Supported by Consejo Nacional de Ciencia y Tecnología - CONACYT (the Mexican National Council for Science and Technology) for the financial support under the grant no. 580617/411550 and registration no. 214885}}
\titlerunning{FunSeqSet}

\author{Juan Carlos S\'{a}enz-Carrasco}
\authorrunning{J. S\'{a}enz-Carrasco}
\institute{The University of Sheffield, S10 2TN UK\\ \email{jcsaenzcarrasco1@sheffield.ac.uk}}

\maketitle              

\begin{abstract}
Dynamic trees, originally described by Sleator and Tarjan, have been studied deeply for non persistent structures providing $\mathcal{O}(\log n)$ time for update and lookup operations as shown in theory and practice by Werneck. However, discussions on how the most common dynamic trees operations (i.e. \textit{link} and \textit{cut}) are computed over a purely functional data structure have not been studied. Even more, asking whether vertices $u$ and $v$ are connected (i.e. within the same forest) assumes that corresponding indices or locations for $u$ and $v$ are taken for granted in most of the literature, and not performed as part of the whole computation for such a question.
We present \seqset, based on the primitive version of \emph{finger tree}s, i.e. the de facto sequence data structure for the purely functional programming language Haskell, augmented with variants of the collection (i.e. sets) data structures in order to manage efficiently $k$-ary trees for the linearisation case of the dynamic trees problem. Different implementations are discussed, and the performance is measured. 

\end{abstract}

\section{Introduction}
\label{Ch-Intro}

A dynamic tree allows three kinds of (basic) operations :
\begin{itemize}
\item Insert an edge.
\item Delete an edge.
\item Answer a question related to the maintained forest property.
\end{itemize}
The first two types of operations are called \emph{updates} and the last one is a \emph{query}. In the simplest case, this is a global question like ``Are vertex $u$ and vertex $v$ in the same tree?" or ``Is vertex $v$ on vertex $u$'s path towards the root?", and the answer is just ``True" or ``False". The purpose of a dynamic tree algorithm is to maintain a forest property faster than by recomputing it from scratch every time the forest changes. The term \emph{dynamic tree problem} was coined by Sleator and Tarjan in \cite{DS-DynTs}. The aims, implementational issues and data structure design by Sleator and Tarjan followed the imperative programming paradigm. We focus our attention in the aforementioned operations under the approach of purely functional programming considering a forest of fixed $n$ number of vertices and consider only undirected edges through this document.

An update in the forest is local. If due to the application the forest changes globally, we can model this with several updates as in performing an unbound sequence of operations over such a forest. In the worst case, we could move from one forest to a totally different one as in a random forest generation. Therefore, it does not make sense to maintain the forest property of the new forest by means of data collected from the old forest faster than by recomputing it with a static forest algorithm. This scenario is suitable for designing and analysing persistent data structures. %Among all types of persistence in data structures we focus on Okasaki's version of the persistent data structure \cite{OkasakiC-PhD}.

Data structures which allow queries and insertion of edges, but not deletion of edges are called \emph{incremental} and \emph{decremental} otherwise \cite{DynTs}. In any case, we can refer to any of the above structures to be \emph{semi-dynamic}. If we want to distinguish between semi-dynamic data structures and data structures allowing both operations, then the latter are called fully \textit{dynamic data structures}. We provide implementation and experimental analysis for the semi and fully dynamic cases.

Note that the term \emph{forest property} is quite general. A forest property can be a predicate on the whole forest (e.g., testing membership), a predicate on pairs of nodes (e.g., connectivity), or a unique component (e.g., the minimum spanning tree, if the weights are such that it is unique). 

The forest property we will mainly deal with in this paper is connectivity. Two vertices $u$ and $v$ are connected, if both vertices are members of the same component or tree. We want to be able to quickly answer each question of the type ``Are vertices $u$ and $v$ connected in the current forest?". 

The paper is organized as follows. Section 2 explains the motivation behind \seqset\ as purely functional data structure. Fundamental structures and mathematical background are explained in Section 3. The actual implementation of \seqset\ is detailed in Section 4, while the experimental analysis applied to it is left to Section 5. Finally, in Section 6, we give our conclusions and describe some topics for future research.

\section{Motivation}
\label{Sec-Motivation}
Inserting and deleting edges are among the most fundamental and also most commonly encountered operations in trees, especially in the dynamic setting. In this paper we deal with trees of degree $n$ and not necessarily rooted or with a specific shape. In \cite{WerneckR-PhD}, Werneck gives a thorough explanation and classification for such a trees and the performance for the aforementioned operations of insertion (i.e. link) and deletion (i.e. cut). For all the cases, the running time is $\mathcal{O}(\log n)$ per operation where $n$ is the number of vertices.

This encourages simplicity and efficiency at the time of the computation so any application can use them. In this section, we motivate the approach of functional programming for these angles. 

\subsection{Applications where dynamic trees operations take place}

Since the definition of the \emph{dynamic trees problem} data structure by Sleator and Tarjan \cite{DS-DynTs}, two major structural operations arise: \emph{link} and \emph{cut}, therefore the term \emph{Link-Cut} trees for this data structure. Besides obvious applications like \textsc{Union-Split-Find} problems \cite{LaiK-ME}, dynamic trees computations are frequently needed in a wide spectrum of applications, to name a few:
\begin{itemize}
\item Flows on Networks; (\cite{LittleBook}, \cite{VertexConn-Nanongkai}) link and cut operations are used to maintain the residual capacities of edges and that of changing labels in the network.
\item Rearrangement of Labelled Trees; recently applied to the problem of comparing trees representing the evolutionary histories of cancerous tumors. Bernardini et al. (\cite{RearrangeLabelledTs}) analyse two updating operations: \emph{link-and-cut} and \emph{permutation}. The former is due to transform the topology of the input trees whereas the latter operation updates the labels without mutating its topology.
\item Geomorphology; Ophelders et al. \cite{Geomorphology} models the evolution of channel networks. Linking and cutting trees are used to model the dynamic behaviour of the growth and shrunk of areas in a river bed.
%\item Machine Learning; recently, (\tcr{citation needed}) \textbf{PyTorch} \cite{PyTorch} (\tcr{description needed}) has emerged thanks by exploiting the use of dynamic graphs (\tcr{description needed}), in comparison with its counterpart %\textbf{TensorFlow}, which makes heavily use of static graphs. Then, dynamic graphs is supported by dynamic trees operations for some computations such maintaining its minimum spanning forest (or tree)
\end{itemize}

\subsection{Dynamic trees in Functional Programming}
Literature has shown a lot about updating edges in trees and graphs, see for instance the handbook for data structures regarding this topic in \cite{DynTs}, but practically little work has been done for the functional programming, specifically for the dynamic setting. 

Some efforts have been done in the real of functional programming. In the case of graph structures, Erwig \cite{InductiveGs} introduces a functional representation of graphs where a graph is defined by induction. Although an interface and some applications have been provided, none of these refer to the dynamic trees problem. For the case of trees, Kmett \cite{LinkCut-Kmett} defines a functional programming version (i.e. in Haskell) of that of the one defined by Sleator and Tarjan \cite{DS-DynTs}; unfortunately Kmett's work relies completely on monads and stateful computation making difficult to reason about the operations and its potential parallelization. Also, the element of a forest is missing in Kmett's work.

\section{Fundamentals}
\label{Ch-Fundamentals}
In this paper we will encounter three different kinds of trees. The first kind of tree is the \emph{input tree} \index{input tree} (i.e. multiway or rose tree \index{Rose tree}), mostly generated by an application such as a graph or a network. The input tree is defined in library \haskell{Data.Tree}\footnote{\url{https://hackage.haskell.org/package/containers-0.2.0.0/docs/Data-Tree.html}} as 
\begin{lstlisting} 
data Tree a  =  Node { rootLabel :: a    
                     , subForest :: Forest a  }
\end{lstlisting} 
where \haskell{Forest a} is defined as 
\begin{lstlisting}
type Forest a = [Tree a]
\end{lstlisting}
The curly brackets refer to the Haskell record syntax, where fields \haskell{rootLabel} and \haskell{subForest} are actually functions to extract the corresponding data out of the \haskell{Tree} type. Since the structure does not constrain the type \haskell{a} to be ordered nor offer any kind of balancing, its performance is linear, that is, querying and updating an element requires to traverse the entire structure to identify the corresponding place in the tree for the operation. So, inserting, deleting or looking up for an element in this kind of tree might take $\mathcal{O}(n)$ per operation.

The following kind of tree is the \emph{set} (i.e. a binary search tree or BST) for query purposes. The common set structure found at Haskell's library repository is \haskell{Data.Set}\footnote{\url{https://hackage.haskell.org/package/containers-0.4.2.0/docs/Data-Set.html}} and defined as
\begin{lstlisting} 
data Set a = Tip 
           | Bin !Size !a !(Set a) !(Set a) 
\end{lstlisting} 
where \haskell{Size} is a synonym of the integer type \haskell{Int}. The exclamation mark refers to a bang annotation, which means that each type next to it will be evaluated to weak-head normal form (for instance by pattern matching on it) or in a strict manner. Within this kind of trees we shall incorporate another similar structure, \haskell{Data.Edison.Coll.LazyPairingHeap}\footnote{\url{http://hackage.haskell.org/package/EdisonCore-1.3.2.1/docs/src/Data-Edison-Coll-LazyPairingHeap.html}}, where its evaluation is lazy when a node holds a single element and partially strict when a node is complete. This one is defined as  
\begin{lstlisting} 
data Heap a = E
            | H1 a (Heap a)
            | H2 a !(Heap a) (Heap a)
\end{lstlisting} 

In theory, both \emph{set} alike structures perform the look up, insertion and deletion in logarithmic time, as this structure is balanced  subject to  elements being constrained to the type class \haskell{Ord}. This is proved in \cite{DataSet} and in \cite{EdisonOverview} for \haskell{Data.Set} and \haskell{Data.Edison.Coll.LazyPairingHeap} respectively.

 In practice, we shall see the \haskell{Data.Edison.Coll.LazyPairingHeap} overtakes \haskell{Data.Set} by a constant factor. In both cases, the common set operations are named the same, that is, \haskell{member}, \haskell{union}, \haskell{insert} share the Haskell function names.

Finally, the third kind of tree is the \emph{finger tree}, i.e. \FT\ which
\begin{itemize}
\item allocates sequences at its leaves, in the form Euler-tour trees as result of \emph{flattening} input trees
\item computes set operations at its intermediate nodes, in the form of monoidal annotations
\end{itemize}

A \FT\ is defined as
\begin{lstlisting}
data FingerTree v a 
   = Empty 
   | Single a 
   | Deep !v !(Digit a) (FingerTree v (Node v a)) !(Digit a)
\end{lstlisting}
where  \haskell{v} is the type of the monoidal annotation, the \emph{set} in our case, and \haskell{a} is the type of the elements in the \emph{sequence}. \haskell{Node} is a type defined below to hold subtrees.
\begin{lstlisting}
data Node v a = Node2 !v a a | Node3 !v a a a 
\end{lstlisting}
Finally, \haskell{Digit} is the type for holding the prefixes and suffixes of the \FT\ tree.
\begin{lstlisting}
data Digit a = One a | Two a a | Three a a a | Four a a a a
\end{lstlisting}

\begin{figure}
\begin{center}
\includegraphics[scale=0.25]{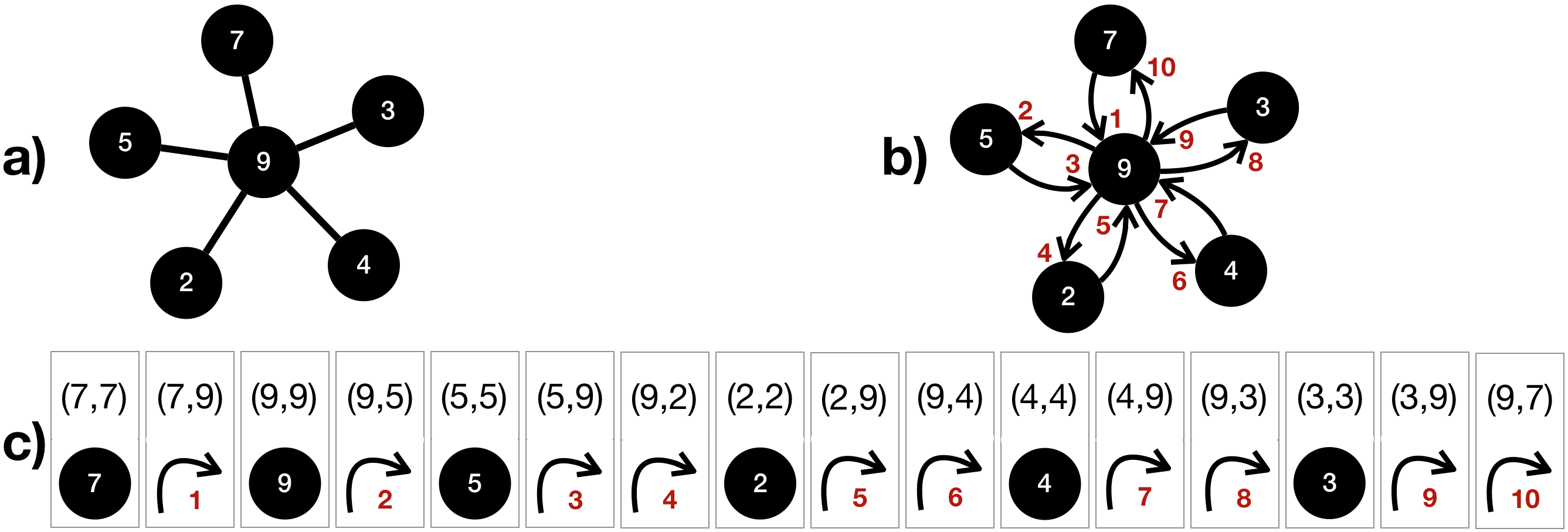} 
\end{center}
\caption{\textbf{a)} input tree with six vertices; \textbf{b)} input tree with broken edges to create a tour rooted at vertex 7 (ETT); and \textbf{c)} the sequence corresponding \textbf{b)}}
\label{fig-anyT}
\end{figure}

\begin{figure}
\begin{center}
\includegraphics[scale=0.25]{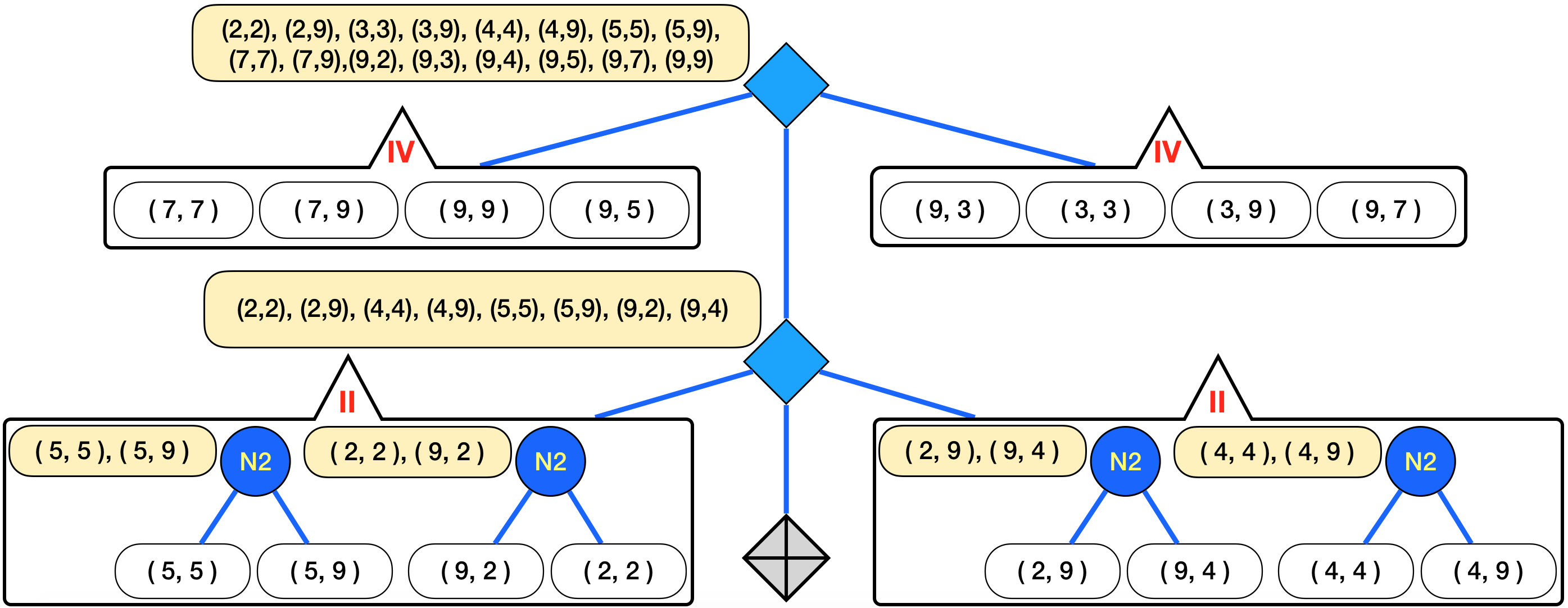} 
\end{center}
\caption{finger tree \FT\ corresponding to the tree in Figure \ref{fig-anyT}, specifically storing the sequence of \textbf{c)}}
\label{fig-FTanyT}
\end{figure}

The following figures depict an example of a finger tree \FT\ of an input tree comprised of six vertices assuming vertex labelled \haskell{7} is the root. In Figure~\ref{fig-anyT} we show an input tree and its corresponding Euler tour tree (ETT) as sequence. In Figure~\ref{fig-FTanyT} we show the corresponding finger tree. Notice that diamond shapes regard the type \haskell{FingerTree}, the white-filled ellipses with single pairs on them correspond to type \haskell{a} and the yellow-filled ellipses holding any number of pairs represent the \emph{sets} for the type \haskell{v}; the rectangular shapes with triangles on top and red Roman numerals are regarded to \haskell{Digit} types and finally, the blue circles represent the \haskell{Node} types to hold subtrees.

The above definition is the general version found in Hinze's and Paterson \cite{FTs} paper, but its implementation can be found in at least two libraries, the default Haskell's sequence \haskell{Data.Sequence}\footnote{https://hackage.haskell.org/package/containers-0.5.0.0/docs/Data-Sequence.html} and as the general-purpose finger tree \haskell{Data.FingerTree}\footnote{\url{http://hackage.haskell.org/package/fingertree-0.1.4.2/docs/Data-FingerTree.html}}, where the former is a specialisation of the latter.

According to Hinze and Paterson, a look up and an update take $\mathcal{O}(\log n)$ amortised per operation, where $n$ is the number of elements in the trees involved in a particular operation. The following is a summary, listed in Table~\ref{tab-FTops}, of the operations we shall use in the paper.
%\begin{center}
\begin{table}
\centering
\begin{tabular}{|| c | l | c ||} 
\hline
Function & Description & Complexity \\  \hline\hline
 \texttt{viewl}       & view the first element    & $\mathcal{O}(1)$ \\ \hline
 $\triangleleft$     & inserting from the left    & $\mathcal{O}(1)$ \\ \hline
 $\triangleright$   & inserting from the right  & $\mathcal{O}(1)$ \\ \hline
 $\bowtie$           & appending two trees (concatenation)  & $\mathcal{O}(\log(min(n_1,n_2)))$ \\ \hline
 \texttt{search}    & looking for an element & $\mathcal{O}(\log n)$ \\ \hline   
\end{tabular}
\caption{In each case, $n$ gives the number of vertices in the first (or only) tree operated upon; for those functions taking two trees as input, $m$ is the number of vertices in the second tree. The result for $\bowtie$ assumes that $m \leq n$ (if not, we can swap the order of the arguments before applying $\bowtie$)}
\label{tab-FTops}
\end{table}
%\end{center}

% The Haskell type signature for the above operations are enlisted below.

\begin{table}
\centering
\begin{tabular}{|| c | l | c ||} 
\hline
Function & Haskell Signature Types\\  \hline\hline
 \texttt{viewl}       &  \haskell{ Measured v a => FingerTree v a -> ViewL (FingerTree v) a } \\ \hline
 $\triangleleft$     &  \haskell{ Measured v a => a -> FingerTree v a -> FingerTree v a } \\ \hline
 $\triangleright$   &  \haskell{ Measured v a => FingerTree v a -> a -> FingerTree v a } \\ \hline
 $\bowtie$           & \haskell{ Measured v a =>  FingerTree v a -> FingerTree v a -> FingerTree v a }  \\ \hline
 \texttt{search}    & \haskell{ Measured v a =>  (v -> v -> Bool) -> FingerTree v a -> SearchResult v a } \\ \hline   
\end{tabular}
\caption{The signatures of the \haskell{FingerTree} operations we call from \seqset }
\label{tab-signatures}
\end{table}

\section{Implementation}
\label{Ch-ETTs}

In this Section, we define our data type for managing trees only supported by three auxiliary functions, \haskell{root}, \haskell{reroot}, and \haskell{pairIn}. We then follow the procedures provided by Tarjan in \cite{DynTsETT} for linking and cutting trees as isolated entities rather than elements of a forest. Finally, we build up the remaining functions and types for the forest structure. For the running time analysis we shall follow the bounds listed in Table~\ref{tab-FTops} as reference and discard the bounds from the set-alike monoidal annotations for two reasons:
\begin{itemize}
\item There is a correlation between the results from benchmarking in Section \ref{sec:Eval} and the performance bounds of those in Table~\ref{tab-FTops}
\item Performance of the set-alike annotation relies on the implementation selected as shown in Figure~\ref{fig:plotSets}
\end{itemize}
So, for instance, inserting an element into a \FT\ from the left ($\triangleleft$) takes $\mathcal{O}(1)$ provided the monoidal annotation performs in $\mathcal{O}(1)$; when the monoidal annotation operation is a set-insertion (e.g., \haskell{Data.Set}) the overall operation performance is then $\mathcal{O}(1 \times \log n) = \mathcal{O}(\log n)$. Now, by observing the plot in Figure~\ref{fig:incLink} we notice that the performance per insertion operation (\haskell{link}) is in practice nearly $\mathcal{O}(1)$.

\subsection{Data types for trees}  
In order to manage a sequence of pairs (representing an Euler-tour tree) we define pairs of type \haskell{a} as \haskell{(a,a)} on the leaves of a finger tree \FT\ where the inner nodes are actual \emph{sets} of type \haskell{Set (a,a)} to support searching within the sequence. We shall denote a prefix \haskell{S.} on sets to refer that data types or functions next to the dot belong to library \haskell{Data.Set} imported as \haskell{S}. We then define such a structure as follows, with its corresponding initial element, the empty tree
\begin{lstlisting}
type TreeEF    a = FingerTree  (S.Set (a,a)) (a,a)

emptyTree :: Ord a => TreeEF a 
emptyTree  = FT.empty 
\end{lstlisting}

\subsection{Helper functions}
We consider the first pair within an Euler-tour tree as the root of a tree, that is
\begin{lstlisting}
root :: Ord a => TreeEF a -> Maybe a  
root tree = case viewl tree of
  EmptyL   -> Nothing
  x :< _   -> Just ( fst x )
\end{lstlisting}
Since \haskell{viewl} takes constant time, \haskell{root} also returns the successful vertex or \haskell{Nothing} in constant time since we just pattern match on its data constructors \haskell{EmptyL} and \haskell{(:<)}. 

When linking two trees, $t_u$ and $t_v$, we consider $t_u$ as a rooted tree at vertex $u$ prior to the insertion of a new edge $(u,v)$. The following is the snippet for such a function, called \haskell{reroot}. 
\begin{lstlisting}
reroot :: Ord a => TreeEF a -> a -> TreeEF a 
reroot tree vertex = case (FT.search pred tree) of
   Position left _ right -> root <| (right >< left)
   _                     -> tree
 where root          = (vertex,vertex)
       pred before _ = (S.member root) before
\end{lstlisting}
Recall that prefixes with a dot mean that the following functions or types are members of the predefined library through its identifier. In the case of \haskell{FT.search}, \haskell{search} is a function imported from \haskell{Data.FingerTree} through the prefix \haskell{FT}. In particular, \haskell{search} returns the data constructor \haskell{Position} following its type, according to Table~\ref{tab-signatures}. The underscore \haskell{_}, one line later, is a wild card. It works as a guard similar to the keyword \haskell{otherwise}, which means that any result from \haskell{search} other than \haskell{Position} will lead to \haskell{tree}, that is, the original tree passed as argument to \haskell{reroot}.
So, rerooting a tree $t$ at vertex $v$ is either $t$, when $v$ is not in $t$, or the pair $(v,v)$ inserted from the left (i.e. the very first element in the sequence) to the concatenation of the right and left subtrees when splitting $t$ at $v$. Since \haskell{reroot} involves one \haskell{search}, one $\triangleleft$ and one $\bowtie$ (i.e. $\mathcal{O}(\log n) + \mathcal{O}(1) + \mathcal{O}(\log n)$), its performance is $\mathcal{O}(\log n)$. 

Testing whether a pair (i.e. edge or vertex) belongs to a tree (i.e. \FT) requires only a boolean answer rather than splitting such a tree, that is,
\begin{lstlisting}
pairIn :: (Measured (S.Set a) a, Ord a)
       => a -> FingerTree (S.Set a) a
       -> Bool
pairIn p monFT = case (FT.search pred monFT) of
  Position _ _ _ -> True 
  _              -> False
 where
   pred before _ = (S.member p) before 
\end{lstlisting}
evaluates the case when the pair \haskell{p} (either a vertex or an edge) is in the given \FT . As soon as \haskell{p} if found or the bottom of the \FT\ is reached, a boolean value is returned. Since a single \haskell{search} is called, \haskell{pairIn} takes $\mathcal{O}(\log n)$.

\subsection{Main functions}
We show that the procedures \emph{link\{v,w\}} and \emph{cut\{v,w\}} by Tarjan in \cite{DynTsETT} regarding Euler-tour Trees (ETT for short) can be implemented declaratively.

We start with the \emph{link} operation:
\begin{displayquote}
Specifically, suppose $link(\{v,w\})$ is selected. Let $T_1$ and $T_2$ be the trees containing $v$ and $w$ respectively, and let $L_1$ and $L_2$ be the lists representing $T_1$ and $T_2$. We split $L_1$ just after $(v,v)$, into lists $L_1^1,L_1^2$, and we split $L_2$ just after $(w,w)$ into $L_2^1,L_2^2$. Then we form the list representing the combined tree by catenating the six lists $L_1^2,L_1^1,[(v,w)],L_2^2,L_2^1,[(w,v)]$ in order. Thus linking takes two splits and five catenations; two of the latter are the special case of catenation with singleton lists.
\end{displayquote}

We call this procedure \haskell{linkTree} and defined like 
\begin{lstlisting} 
linkTree :: Ord a => a -> TreeEF a -> a -> TreeEF a -> Maybe (TreeEF a) 
linkTree u tu v tv = case (pairIn (u,u) tu, pairIn (v,v) tv) of
  (False, _    ) -> Nothing
  (_    , False) -> Nothing 
  (True , True ) -> Just £
    let from = reroot tu u
        (Position left _ right) = FT.search pred tv
    in  ((left >| (v,v)) >| (v,u)) >< from >< ((u,v) <| right)
 where
   pred before _ = (S.member (v,v)) before
\end{lstlisting}

The first four lines confirm that the vertices \haskell{u} and \haskell{v} belong to their corresponding trees \haskell{tu} and \haskell{tv}. The tree \haskell{from} is transformed in such a way that left vertex (\haskell{u}) is now the root. Now, by \haskell{search}ing vertex \haskell{v} in tree \haskell{tv} (i.e. \haskell{FT.search pred tv}) we are actually splitting tree \haskell{tv} into subtrees \haskell{left} and \haskell{right}. Now, the remaining task is simply gluing all subtrees with the new edges in order. Thus, in our function we required two splits (one local and one from \haskell{reroot}) and three catenations ($\bowtie$: two local and one from \haskell{reroot}) rather than five from Tarjan's procedure. Like Tarjan's procedure, our function \haskell{linkTree} guarantees termination as it is not recursive, hence is computed in a $\mathcal{O}(1)$ number of steps. Unlike Tarjan's procedure, our function is not only the declarative specification but the actual computation of the \emph{link} operation of dynamic trees problem with a reduced number of concatenations, that is, $\mathcal{O}(\log n)$ amortised.

Now, for the \emph{cut} operation, following Tarjan's definition we have:
\begin{displayquote}
Similarly, suppose we wish to perform $cut(\{v,w\})$. Let $T$ be the tree containing $\{v,w\}$, represented by list $L$. We split $L$ before and after $(v,w)$ and $(w,v)$, into $L^1$, $[(v,w)]$, $L^2$, $[(w,v)]$, $L^3$ (or symmetrically $L^1$, $[(w,v)]$, $L^2$, $[(v,w)]$, $L^3$). The lists representing the two trees formed by the cut are $L^2$ and the list formed by catenating $L^1$ and $L^3$. Thus cutting takes four splits (of which two are the special case of splitting off one element) and one catenation. 
\end{displayquote}

We call the above procedure \haskell{cutTree} and is defined as 
\begin{lstlisting}
cutTree :: Ord a => a -> a -> TreeEF a -> Maybe (TreeEF a,TreeEF a) 
cutTree u v tree = case FT.search predUV tree of
 Position left _ right ->
   case (FT.search predVU left ) of
      Position leftL _ rightL ->           -- (v,u) is on the left 
        Just (rightL, leftL >< right)
      _              ->                    -- (v,u) is on the right
        case (FT.search predVU right) of
          Position leftR _ rightR ->
            Just (leftR, left >< rightR)
          _ -> Nothing           -- BAD Formed tree since (v,u) is missing 
 _  -> Nothing            -- BAD Formed tree since (u,v) is missing     
 where
   predUV before _ = (S.member (u,v)) before 
   predVU before _ = (S.member (v,u)) before 
\end{lstlisting} 
With the help of case analyses, we verify that input edge $(u,v)$ (and its corresponding $(v,u)$) is a member of the input tree \haskell{tree}. If that is not the case, \haskell{Nothing} is return, meaning the input edge is not in \haskell{tree}. Otherwise, like in Tarjan's procedure, we compute one catenation and unlike Tarjan, we perform at most three splits, that is, in the worst case we have $1 \times \mathcal{O}(\bowtie) + 3 \times \mathcal{O}($\haskell{search}$)$, that is, $\mathcal{O}(\log n)$ amortised per \haskell{cutTree} operation.

\subsection{Managing forests}
Recalling the \emph{forest property} within the context of the dynamic trees problem, we are at the point to define the structure that holds everything in place, 
\begin{lstlisting}
type ForestEF  a = FingerTree  (S.Set (a,a)) (TreeEF  a) 

emptyForest :: Ord a => ForestEF a  
emptyForest  = FT.empty 
\end{lstlisting}

We take advantage once again of the \FT\ benefits about updates and look ups of, in this case, trees as atomic elements. Prior to performing the update operations within a forest, we define the lookup ones.

When looking for a single vertex $v$ (i.e. the pair \haskell{(v,v)}), we simply apply the \haskell{search} function from \haskell{Data.FingerTree} to the forest \haskell{f} provided as second argument. Recall, from its Haskell type, that the simplest element in \haskell{f} is a tree (i.e. an ETT). Then, the successful search on a \FT\ returns three elements: left subtree, searched element, and the right subtree. In this case, the left and right subtrees are actually subforests which are discarded. We then return \haskell{tree}, the tree containing vertex $v$ and its root wrapped as \haskell{Maybe} type. In case of unsuccessful search, we simply return \haskell{Nothing}.
\begin{lstlisting}
searchFor :: Ord a => a -> ForestEF a -> Maybe (TreeEF a, a) 
searchFor v f = 
 case FT.search pred f of 
  Position _ tree _ -> Just (tree, fromJust (root tree) ) 
  _                 -> Nothing
 where
   pred before _ = (S.member (v,v)) before 
\end{lstlisting}
Since we just apply \haskell{search} once, this function takes $\mathcal{O}(\log n)$.

Now, for the claimed forest property for dynamic trees problem, we define the function \haskell{connected} as the following case analysis. We input two vertices \haskell{x} and \haskell{y}, as an edge, and a forest \haskell{f}. If the search of any of the vertices \haskell{x} or \haskell{y} on \haskell{f} is unsuccessful we return \haskell{Nothing}, otherwise we test equality on the roots of the returning trees. If both roots turn out to be the same then claim that edge $\{x,y\} \in f$, specifically $\{x,y\}$ is in the same component; on the other hand, if roots are different we return the corresponding trees (and their roots) altogether with \haskell{False} as an answer to the connectivity question within the forest.
\begin{lstlisting}
type PairTreeVertex a = (TreeEF a, a, TreeEF a, a) 

connected :: Ord a => a -> a -> ForestEF a -> (Bool, Maybe (PairTreeVertex a)) 
connected x y f = 
 case (searchFor x f, searchFor y f) of 
  (Nothing          , _           ) -> (False, Nothing) 
  (_                , Nothing     ) -> (False, Nothing) 
  (Just (tx,rx)     , Just (ty,ry)) -> if rx == ry 
                                   then (True,  Just(tx,rx,tx,rx))  
                                   else (False, Just(tx,rx,ty,ry)) 
\end{lstlisting}
We have applied \haskell{searchFor} twice, hence \haskell{connected} is performed in $\mathcal{O}(\log n)$ amortised.

We proceed to define the update operations over a forest. In the case of a \emph{link}, we firstly pattern match the trivial case (i.e. input vertices are the same) and cases for whether the new edge is already in the same component (i.e. same tree). On an unsuccessful connectivity, we perform \haskell{link} through \haskell{linkTree}, otherwise we return the original forest \haskell{f}. Returning the same input forest does not offer feedback when unsuccessful linking but allows the computation of dynamic operations as a sequence fluently.
\begin{lstlisting}
link :: Ord a => a -> a -> ForestEF a -> ForestEF a 
link x y f 
  | x == y    = f  
  | otherwise = 
     case connected x y f of 
      (False, Just (tx,rx,ty,ry)) -> case (linkTree x tx y ty) of
         Nothing     -> f
         Just result -> linkAll result 
      _                           -> f 
 where 
    Position lf' _ rf' = FT.search predX f 
    Position lf  _ rf  = FT.search predY (lf' >< rf') 
    linkAll tree    = tree <| (lf >< rf)
    predX before _ = (S.member (x,x)) before 
    predY before _ = (S.member (y,y)) before 
\end{lstlisting} 
The overall performance of a \emph{link} operation is two splits and two catenations alongside the performance of \haskell{connected} and \haskell{linkTree}, that is, $2\times \mathcal{O}(\log n) + 2\times \mathcal{O}(\log n) + \mathcal{O}(\log n) + \mathcal{O}(\log n)$. Since this function is static at runtime, the performance for \haskell{link} is $\mathcal{O}(\log n)$ amortised.

Finally, the \haskell{cut} operation over forests. Like \haskell{link}, we case analysis on the trivial case and on connectivity. If the latter is successful we perform \haskell{cutTree}. Again, if cutting a tree is not possible we return the same input forest to allow fluency when applying a sequence of operations over a forest.
\begin{lstlisting}
cut :: Ord a => a -> a -> ForestEF a -> ForestEF a 
cut x y f  
 | x == y    = f  
 | otherwise = 
    case connected x y f of 
      (True, Just (tx,_,_,_)) -> case (cutTree x y tx) of
        Nothing     -> f 
        Just result -> buildForest result  
      _                       -> f 
 where 
    buildForest (t2,t3) = t2 <| (t3 <| (lf >< rf)) 
    Position lf _ rf = FT.search pred f
    pred before _    = (S.member (x,x)) before
\end{lstlisting}
The overall performance of a \emph{cut} operation is one split, one catenation, one connectivity testing and the application of \haskell{cutTree} once, that is, $\mathcal{O}(\log n) + \mathcal{O}(\log n) + \mathcal{O}(\log n) + \mathcal{O}(\log n)$. Since this function is static at runtime, the performance for \haskell{cut} is $\mathcal{O}(\log n)$ amortised.

\section{Experimental Analysis}
\label{sec:Eval} 
This section presents experiments to evaluate how much running time costs in terms of performance. The experiments will show that, in practice, \seqset\ is faster than expected in the theoretical analysis (Section\ref{Ch-ETTs}).

This section is organised as follows. Firstly, we describe the experimental setup. Secondly, a brief description in the implementation of test sets is provided. We then present experimental studies of the three different operations in \seqset . Finally, we present an additional experiment for the cases where laziness as speeding up factor in favour of the running times for the dynamic tree operations.

\subsection{Experimental Setup}
Functions \haskell{linkTree}, \haskell{cutTree}, \haskell{link}, \haskell{cut}, \haskell{connected}, and the ones described as ``helper functions"  were implemented by the author in Haskell and compiled with \haskell{ghc} version 8.0.1 with optimisation \haskell{-O2}. The experiments were performed on a 2.2 GHz Intel Core i7 MacBook Pro with 16 GB 1600 MHz DDR3 running macOS High Sierra version 10.13.1 (17B1003). We imported the following libraries into our code from the online package repository Hackage: \cite{HaskellFT} code for finger trees, \cite{HaskellSet} for conventional sets and \cite{HaskellEdison} for lazy sets.

The running time of a given computation was determined by the mean of three executions.

\subsection{Data structure} 
The values maintained by the data structures (sets and finger trees) are stored as fixed-precision \haskell{Int} types, holding values from $-2^{63}$ up to $2^{63}-1$ although we test only the positive values. The structures are initialized with a fixed number of nodes (or vertices) $n$; this number does not change during the execution. This allows us to know the initial size of the forest and we subtract it from the benchmarking.

Since \seqset\ is not called by any application, the random generation of nodes for \haskell{link} or \haskell{cu}t will not necessarily be effective. Actually, around 70\% of the generated nodes $x$ and $y$ passed to \haskell{link} and \haskell{cut} were not valid, that is, their result turned out to be the original forest. In order to overcome this, we stored the randomly generated nodes that were effective into plain files and from there benchmarking the dynamic tree operations.

\subsection{Incremental operations} 
\begin{figure}
\begin{center}
\includegraphics[scale=0.38]{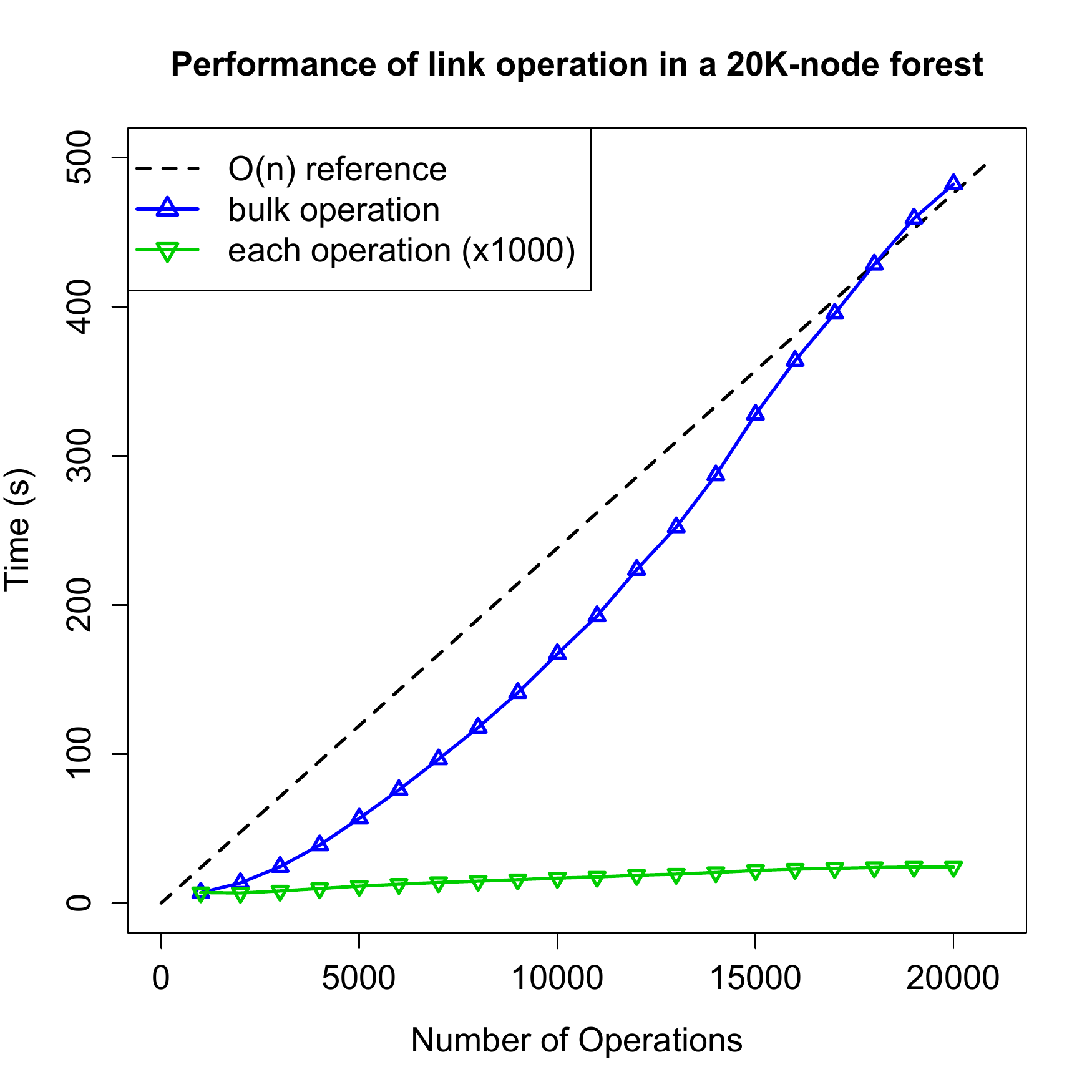} 
\end{center}
\caption{Sequence of {\link}s from empty forest up to a single tree in such forest}
\label{fig:incLink}
\end{figure}

We start with an empty forest (just singleton-trees); given $n=20,000$ nodes we perform $1 \ldots 20,000$ \haskell{link} operations. Upon reaching a target length, we plot the total time taken. Then, we divide the time taken by the number of operations to calculate the time per operation and then multiply it by a constant (x1000) to make the curve visible in the same chart, Fig~\ref{fig:incLink}.

\textit{\emph{Results}}. The behaviour of the curve regarding the time per \haskell{link} operation shows that in practice it takes $\mathcal{O}(1)$, as expected. Same applies to the case when $n$ operations are applied in bulk, that is $\mathcal{O}(n)$.

\subsection{Fully dynamic operations} 
\begin{figure} %[H]
\centering
\begin{subfigure}{.5\textwidth}
  \centering
  \includegraphics[scale=0.38]{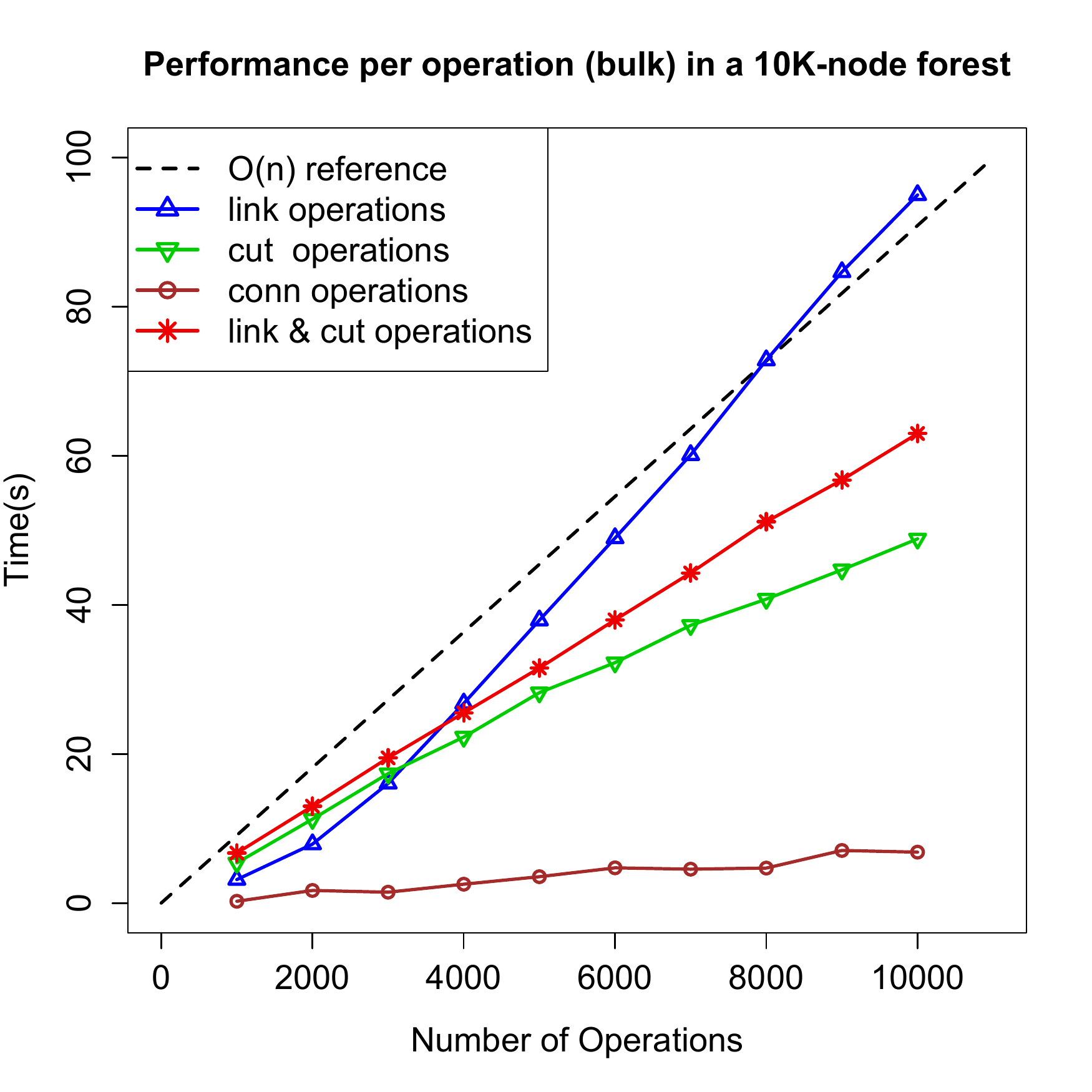}
  \caption{In bulk}
  \label{fig:sub1}
\end{subfigure}%
\begin{subfigure}{.5\textwidth}
  \centering
  \includegraphics[scale=0.38]{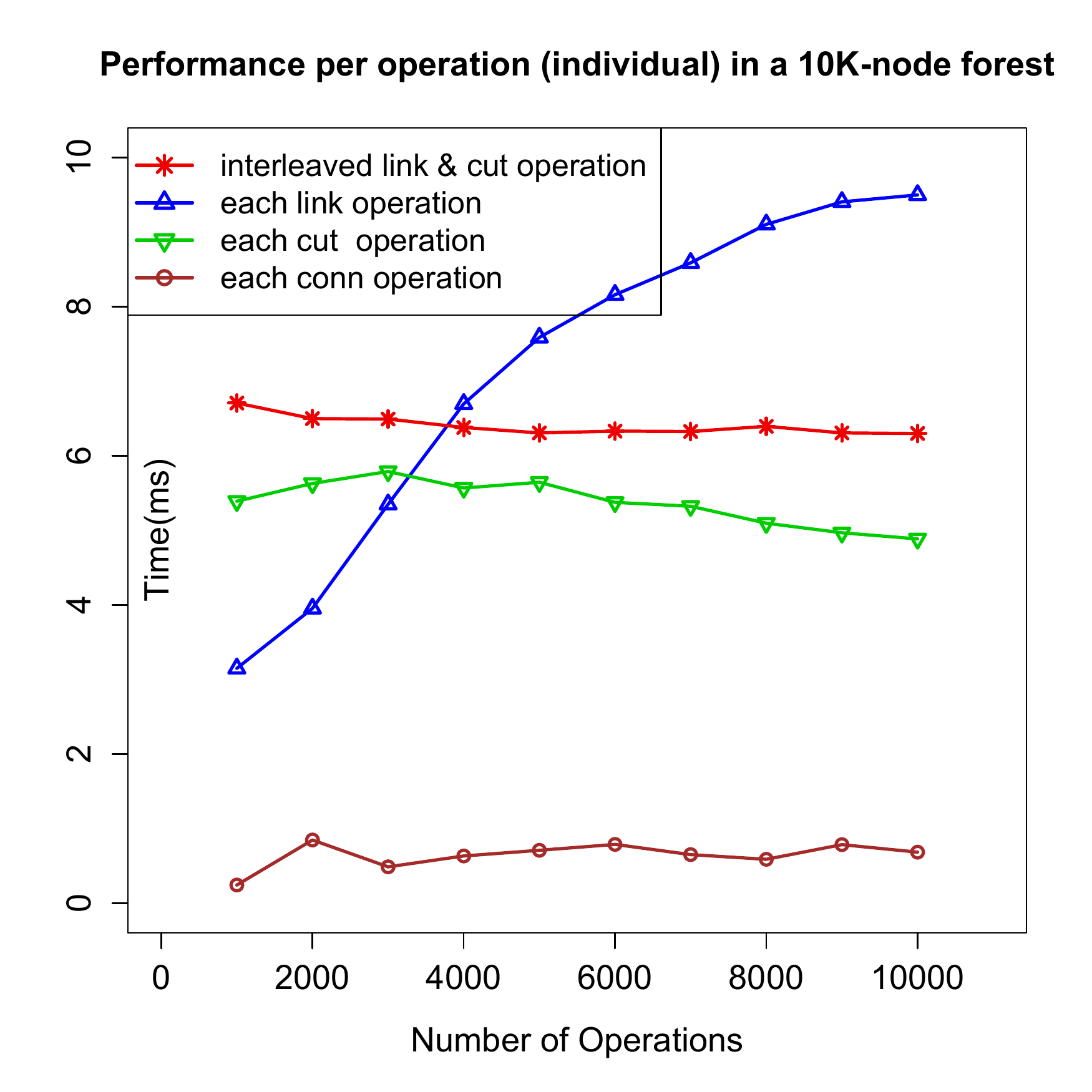}
  \caption{Per operation}
  \label{fig:sub2}
\end{subfigure}
\caption{Time taken by operation} %, and interleaved \haskell{link} and \haskell{cut}}
\label{fig:EachOp}
\end{figure}

We start with the incremental process as before for $n=10,000$. Then, for \haskell{cut} we start in the opposite direction, that is, cutting from a single tree in the forest until only singleton-trees remain in such forest. To this performance we subtract the time taken for the incremental bit. For \haskell{connected} performance we compute first an interleaved operation of \haskell{link} and \haskell{cut} (not necessarily in this order). We measure the time taken for \haskell{connected} followed by the corresponding \haskell{link} or \haskell{cut} and then we subtract the interleaved process. Figures \ref{fig:sub1} and \ref{fig:sub2} show our three dynamic operations in bulk and per operation.

\begin{figure} %[H]
\centering
\begin{subfigure}{.5\textwidth}
  \centering
  \includegraphics[scale=0.38]{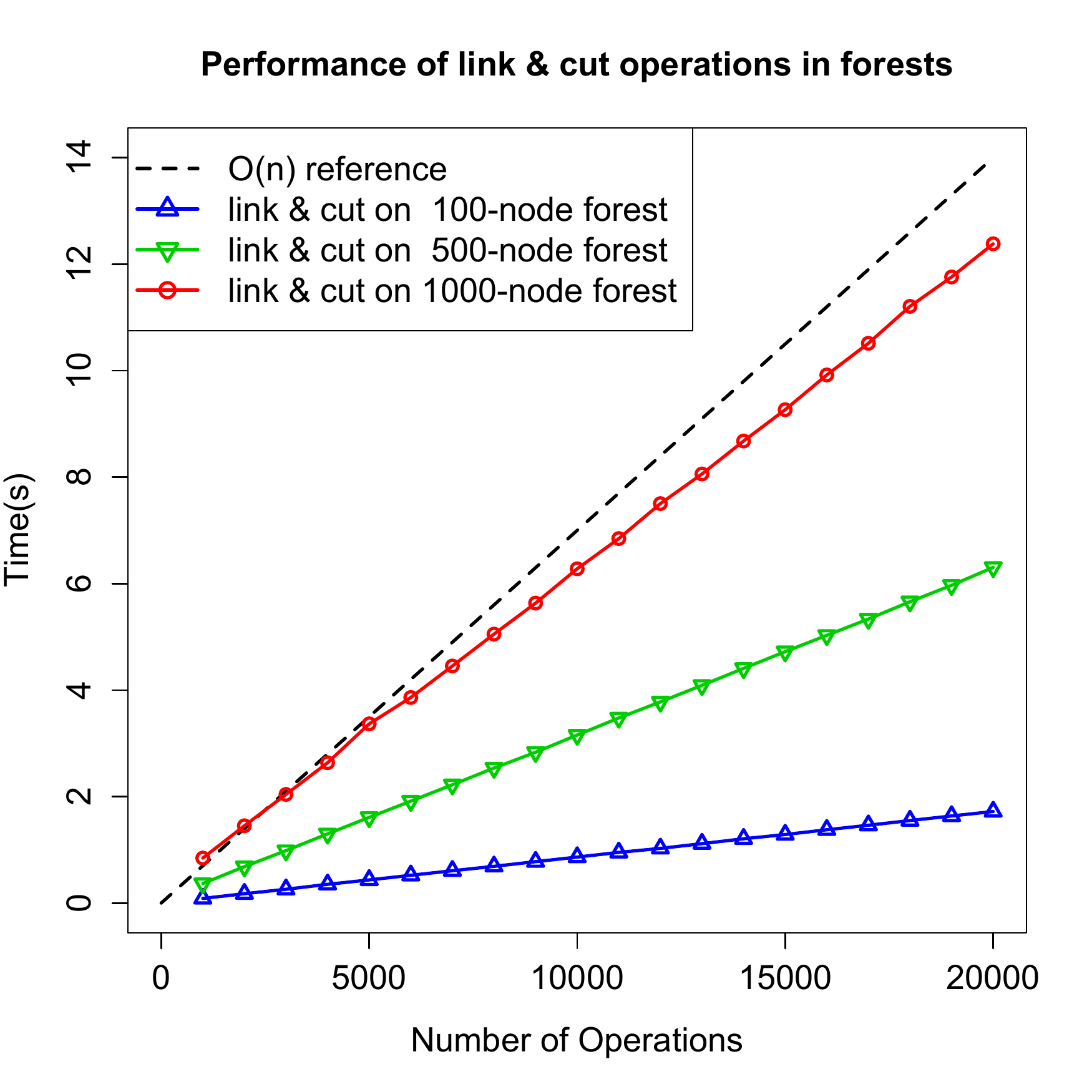}
  \caption{In bulk}
  \label{fig:ssub1}
\end{subfigure}%
\begin{subfigure}{.5\textwidth}
  \centering
  \includegraphics[scale=0.38]{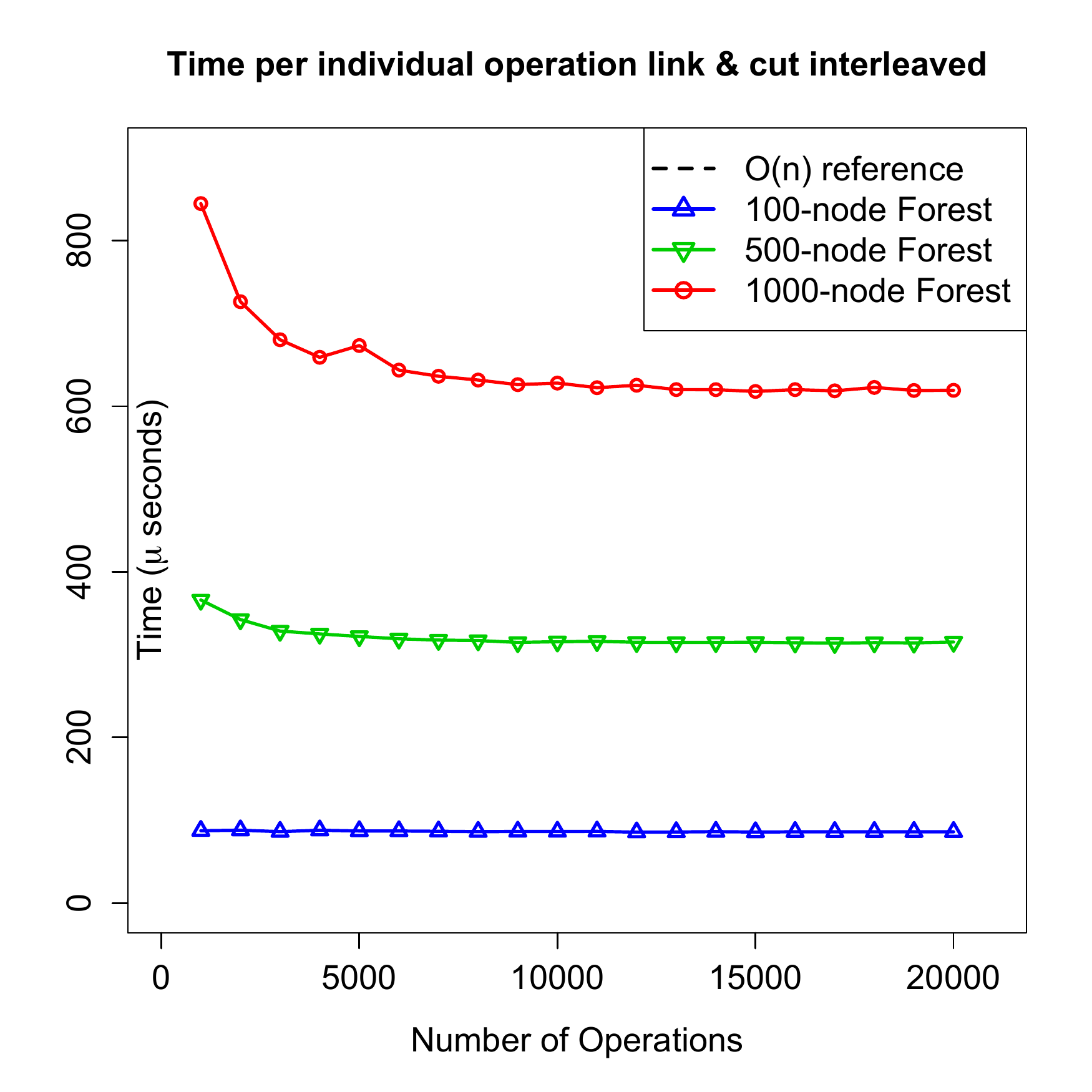}
  \caption{Per operation}
  \label{fig:ssub2}
\end{subfigure}
\caption{Time taken when \haskell{link} and \haskell{cut} are interleaved with different forest sizes}
\label{fig:EachOp}
\end{figure}

\textit{\emph{Results}}. We observe that \haskell{cut} and \haskell{connected} obey the same pattern as \haskell{link}. That is, $\mathcal{O}(1)$ time per operation being \haskell{connected} the fastest of the dynamic tree operations, as expected. From the above analyses, we notice that \haskell{link} performs better when it is interleaved with \haskell{cut}. To see this behaviour closer, we present the bulk and individual cases in the following charts, Figures \ref{fig:ssub1} and \ref{fig:ssub2}, varying the forest size under the same amount of operations.

\subsection{Selection of the set data structure}
\begin{figure}
\begin{center}
\includegraphics[scale=0.4]{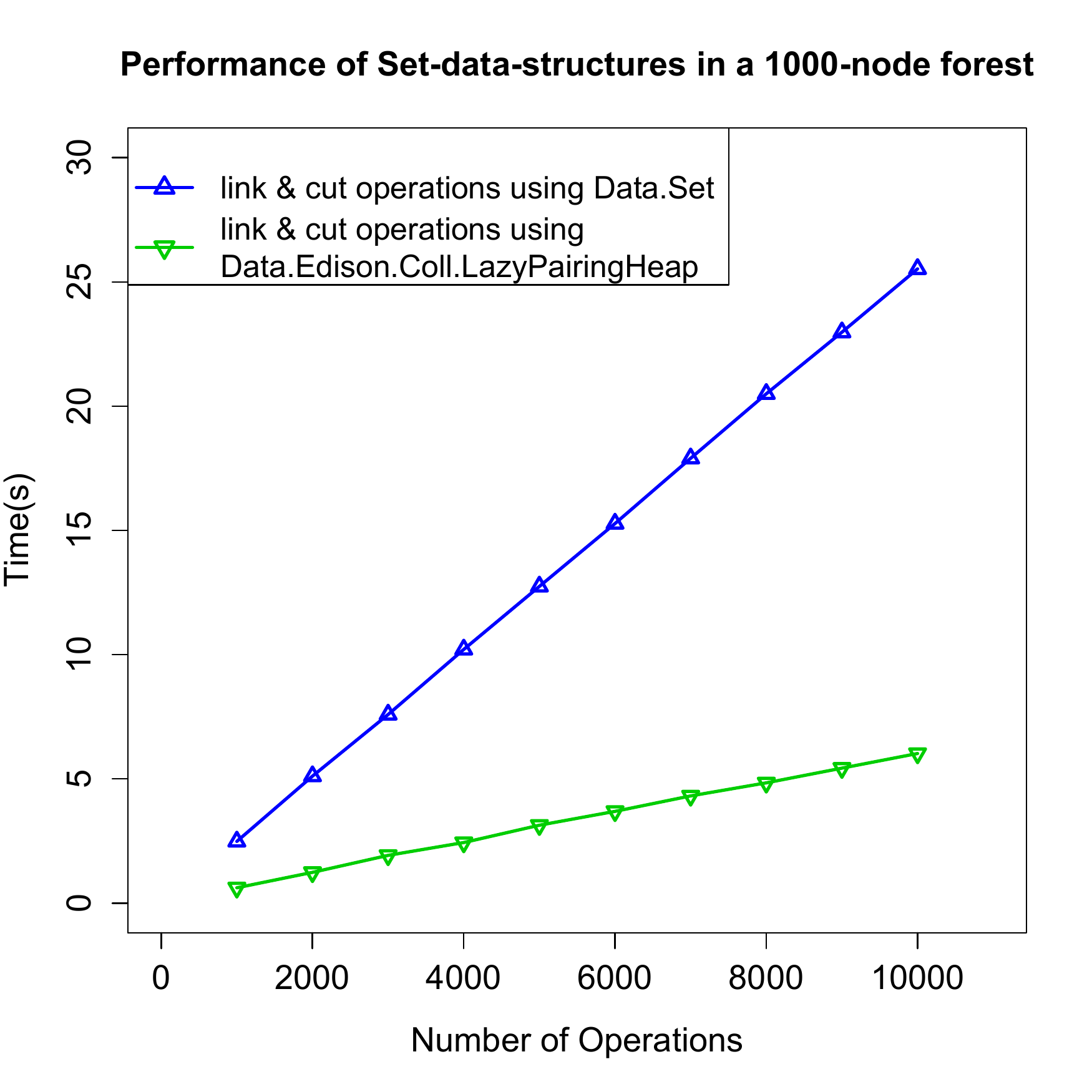} 
\end{center}
\caption{Dynamic operations through different sets structures as monoidal annotations}
\label{fig:plotSets}
\end{figure}

The set-like data structure is crucial in our implementation and testing of \seqset since is the search engine for the nodes when any operation is applied to a forest. There are plenty of implementations for such set-like structure, mostly as binary balanced search trees. In our case, where Haskell is a lazy-evaluation language by default, we select two main choices to compare: \haskell{Data.Set} which is a strict data type definition and \haskell{Data.Edison.Coll.LazyPairingHeap} which is semi-lazy or semi-strict data type. Figure \ref{fig:plotSets} shows the performance for each.

The above curves show that, although by a constant factor, laziness speeds up the running time in the computation of dynamic tree operations through the set-like data structures.

\section{Conclusions and Further Work}
\label{Ch-Conclusion}
\subsection{Final remarks}

For the first time, the purely functional programming approach is discussed for the dynamic trees problem, in particular for the linearisation case. All data structures we discuss above can solve the dynamic connectivity problem for trees maintaining a forest under a finite sequence of edge insertions and deletions and supporting queries asking whether two vertices belong to the same tree or not. We have presented \seqset , a new approach to two existent functional data structures (i.e. finger trees and sets) for maintaining dynamic trees. This structure can manage $k$-degree trees, rooted or unrooted persistently whilst solving the dynamic trees problem. 

Although the updates are conceptually very simple, namely \haskell{link} and \haskell{cut}, the proof that both indeed take $\mathcal{O}(\log n)$ time is rather inherited from the core structures. Such definitions are acyclic and involve $\mathcal{O}(1)$ number of steps to perform each. Our experimental analysis has shown that the three operations we have implemented meet the theoretical bounds of $\mathcal{O}(\log n)$. Also, a native mechanism in the Haskell programming language, the lazy evaluation, is a crucial factor to achieve such performance. Despite the fact that  the \haskell{link} operation is the slowest, it still runs within the $\mathcal{O}(\log n)$ bound.

\subsection{Further work}
Uniqueness on edges allow to carry labels, therefore \seqset\ could solve the dynamic trees problem from other approaches such as path-decomposition (i.e. link-cut trees) and tree-contraction. 

Parallelism can play an important speed up when calling functions such \haskell{connected}. Recalling the its first lines, we have
\begin{lstlisting} [mathescape]
connected x y forest = 
   case (searchFor x forest, searchFor y forest) of 
$\ldots$
\end{lstlisting}
Since the result of the leftist \haskell{searchFor} is independent from the right one, both are suitable candidates to be evaluated in parallel. The pending research here is the time and space analysis between the sequential (both strict and lazy) against the parallel cases.

\bibliography{./refs/refs}
\bibliographystyle{splncs04}

\end{document}